\documentclass[aps,nofootinbib,prd,superscriptaddress,tightenlines,notitlepage,twocolumn,showpacs]{revtex4-1}
\usepackage{amsmath}
\usepackage{mathrsfs}
\usepackage{makecell}
\usepackage{multirow}
\usepackage{graphicx}
\usepackage{bm}
\usepackage[varg]{txfonts}
\usepackage{enumerate}
\usepackage[normalem]{ulem}
\usepackage[OT1]{fontenc}
\usepackage[colorlinks]{hyperref}

\begin{document}

\title{Constraints on Einstein-dilation-Gauss-Bonnet gravity and electric charge of compact binary systems from GW230529}
\author{Bo Gao}
\affiliation{Key Laboratory of Dark Matter and Space Astronomy, Purple Mountain Observatory, Chinese Academy of Sciences, Nanjing 210033, China}
\affiliation{School of Astronomy and Space Science, University of Science and Technology of China, Hefei, Anhui 230026, China}
\author{Shao-Peng Tang}
\affiliation{Key Laboratory of Dark Matter and Space Astronomy, Purple Mountain Observatory, Chinese Academy of Sciences, Nanjing 210033, China}
\author{Hai-Tian Wang}
\email[Corresponding author: ]{wanght@pku.edu.cn}
\affiliation{Kavli Institute for Astronomy and Astrophysics, Peking University, Beijing 100871, China}
\author{Jingzhi Yan}
\email[Corresponding author: ]{jzyan@pmo.ac.cn}
\affiliation{Key Laboratory of Dark Matter and Space Astronomy, Purple Mountain Observatory, Chinese Academy of Sciences, Nanjing 210033, China}
\affiliation{School of Astronomy and Space Science, University of Science and Technology of China, Hefei, Anhui 230026, China}
\author{Yi-Zhong Fan}
\affiliation{Key Laboratory of Dark Matter and Space Astronomy, Purple Mountain Observatory, Chinese Academy of Sciences, Nanjing 210033, China}
\affiliation{School of Astronomy and Space Science, University of Science and Technology of China, Hefei, Anhui 230026, China}
\date{\today}

\begin{abstract}
    In this work, we study the implications of GW230529 on gravity theories and the charge of black holes. The GW230529, which was initially released in O4a, is most likely neutron star-black hole (NSBH) mergers. We reanalyze the data from the GW230529 event to obtain bounds on the Einstein-dilation-Gauss-Bonnet (EdGB) gravity parameter $\sqrt{\alpha_{\rm EdGB}}$ and the electric charge of compact binary systems. The event places a $90\%$ credible upper bounds on $\sqrt{\alpha_{\rm EdGB}}$ of $\lesssim 0.298$ km. After including high order corrections of EdGB gravity, the bounds improve to $\sqrt{\alpha_{\rm EdGB}} \lesssim 0.260$ km. Analyses of GW230529 also yield a $90\%$ credible upper bounds on the combination of charge-to-mass ratio of the binary components $\zeta \lesssim 0.024$. The constraints are more stringent than those derived from previously observed single gravitational wave merger event.
\end{abstract}

\maketitle

\section{Introduction} \label{sec-intro}
The fourth observing run (O4) of the Advanced LIGO (aLIGO), Advanced Virgo, and KAGRA observatory network (LVK) commenced on 24 May 2023. The first part of the fourth observing run (O4a) concluded at 16:00 UTC on 16 January 2024. Shortly after the start of O4a, aLIGO detected a gravitational wave (GW) signals GW230529\_181500 (hereafter referred to as GW230529) with a signal-to-noise ratio (SNR) of $11.6$. GW230529 was emitted from the coalescence of a compact binary star with masses in the ranges of $2.5-4.5 \, {\rm M_\odot}$ and $1.2-2.0 \, {\rm M_\odot}$, as measured at the $90\%$ credible intervals \citep{2024arXiv240404248T}. The analysis of GW230529 does not provide information about the tidal deformability of the secondary object and the primary object's tidal deformability is estimated to peak at zero. Based on the GW data alone, definitively identifying the components of GW230529 is remains challenging. However, by considering existing estimations for the maximum neutron star mass \citep{2024PhRvD.109d3052F}, the most plausible interpretation is that GW230529 resulted from the merger of a neutron star and a black hole. 
Notably, the extended inspiral signals observed by aLIGO in GW230529 could provide more stringent constraints on deviations from general relativity or the presence of charge in black holes within the parameterized post-Einsteinian (ppE) framework \citep{2009PhRvD..80l2003Y,2016PhRvD..94h4002Y}

Scalar Gauss-Bonnet (sGB) gravity extends General Relativity (GR) by incorporating a dynamical scalar field coupled to the Gauss-Bonnet invariant.
This modification introduces a coupling constant $\alpha_{\rm GB}$, which has a dimension of length squared.
The literature on sGB gravity and its implications is extensive and continues to evolve (see, for instance, Refs.~\citep{Nojiri:2010wj,Nojiri:2017ncd} for comprehensive reviews and Refs.~\citep{2018PhRvL.120m1104S, Fernandes:2022zrq,2022CQGra..39w5015F, 2022PhRvD.106d4018E, Nojiri:2023jtf, 2024PhRvD.109d4046N, 2022PhRvD.106d1503C} for recent developments).
A particular variant of sGB gravity, Einstein-dilaton Gauss-Bonnet (EdGB) gravity, introduces a dilaton scalar field, motivated by string theory and cosmological inflationary models \citep{Kanti:1995vq, 1997PhRvD..55..739T}. Significant effort has been devoted to constraining the parameter $\sqrt{\alpha_{\rm EdGB}}$, with a comprehensive summary available in Table~\ref{tab:constr}. In addition to studying constraints from ground-based GW detectors, there are some researches on constraints from future space-based GW detectors. For instance, the TianQin mission is expected to refine these constraints to an order of $\mathcal{O}(0.1)$ km using extreme mass ratio GW sources \cite{2024arXiv240205752T}. Additionally, the capability of space-based GW observatories such as LISA, TianQin, and Taiji to constrain EdGB gravity has been discussed in recent literature \citep{Shi:2022qno,2023ChPhC..47j5101S,Luo:2024vls}. For events involving high-mass systems, the quasinormal mode corrections of EdGB gravity are significant \citep{Carson:2020ter}. GW from extreme mass ratio inspirals have also gained noticeable attention due to their potentiality in testing deviations from Kerr geometry \citep{Rahman:2022fay,AbhishekChowdhuri:2023gvu}. This paper focuses on the corrections in the inspiral phase, as both components of GW230529 are of relatively low mass.
\begin{table*}[ht]
    \caption{Constraints on $\alpha_{\rm EdGB}$ in previous studies} \label{tab:constr}
    \begin{tabular}{ccccccccc}
    \hline \hline
    {} & LMXB & NS &  \multicolumn{2}{c}{GW(BBH)} & \multicolumn{2}{c}{GW(NSBH)} &  \multicolumn{2}{c}{Stacking} \\ \hline
    \multirow{6}{*}{$\sqrt{\alpha_{\rm EdGB}}[{\rm km}]$} & \multirow{6}{*}{1.9 \citep{2012PhRvD..86h1504Y}} & \multirow{6}{*}{1.29 \citep{2021PhRvD.104l4052S}} 
    & 0.27 (0.254)$^a$ & GW190814 \citep{2023PhRvD.108d4061W} & &  & 0.25 (0.98)$^b$ & GWTC-3 \citep{2023PhRvD.108d4061W} \\ 
    && & \multirow{2}{*}{5.6} & \multirow{2}{*}{GW170608 \citep{2019PhRvL.123s1101N}} & \multirow{2}{*}{1.1 (0.87)} & \multirow{2}{*}{GW200115 \citep{2023PhRvD.108d4061W}} & \multirow{2}{*}{1.18} & GW200105, GW200115, \\ 
    &&&&&&&& GW190814 and BBHs \citep{2022PhRvD.105f4001L} \\ 
    && & 4.3 & GW151226 \citep{2019PhRvD.100j4001T} & 1.33 & GW200115 \citep{2022PhRvD.105f4001L} & 1.85 & GWTC-1 \citep{2019PTEP.2019j3E01Y} \\ 
    && & 0.4 & GW190814 \citep{2021PhRvD.104b4015W} & 2.18 & GW190814 \citep{2023PhRvD.108d4061W} & 1.7 & GWTC-1 and GWTC-2 \citep{2021PhRvD.104b4060P} \\
    & & & 0.8 & GW190814 \citep{Carullo:2021dui} &&&& \\ \hline \hline
    \multicolumn{9}{l}{Notes. Values in parentheses represent results from analyses with merge-ringdown contributions.} \\
    \multicolumn{9}{l}{~~~~~~~~~~~~Assuming GW190814 as a NSBH merger, the constraints are 0.98.} \\
    \end{tabular}
\end{table*}

Astrophysical black holes (BHs) are typically assumed to be Kerr black holes due to de-charging processes from their surrounding environments. However, GR \citep{Liu:2020cds, maier2024charged} and theories beyond the Standard Model \citep{Cardoso:2016olt} suggest that BHs can possess various types of charges. Moreover, observational evidence confirming the charge neutrality of BHs remains elusive. Consequently, there is a strong impetus to determine the electric charge of BHs. Several methodologies have been developed to limit the electric charge of BHs. For example, imaging supermassive black holes with the Event Horizon Telescope has provided constraints on the charge-to-mass ratio at a $68\%$ credible level, yielding a value of 0.72 \citep{EventHorizonTelescope:2021dqv, EventHorizonTelescope:2022xqj, Ghosh:2022kit}. Additionally, studies on the bremsstrahlung emission profile of BHs \citep{2018NatAs...2..764K, 2018MNRAS.480.4408Z} have established an observational upper bound on the charge of Sgr $\rm A^{*}$, which $\lesssim 3\times 10^8$ C. The presence of electric charge in BHs could also alter the waveform of GW during the coalescence of binary compact objects. This modification provides a novel avenue to constrain BH charges through GW events. Specifically, the inspiral, merger-ringdown signals, and electromagnetic counterparts associated with GWs can individually provide bounds on the charge of BHs \citep{Wang:2020ori, 2021PhRvL.126d1103B,2021PhRvD.104j4063W,2022PhRvD.105f2009C,2023PhRvD.108h3018Y, 2024PhRvD.109b4058G}.

In this work, by incorporating modifications to the GW phase within the ppE framework, we conduct Bayesian analyses using GW230529 data to constrain parameters in EdGB theories and the potential charge of black holes. The refined results from the inspiral signal of the GW data establish more stringent limits compared to previous studies. The methodology employed, involving the ppE framework and Bayesian analysis, is detailed in Section~\ref{sec-met}. The main findings are discussed in Section~\ref{sec-res}, and a summary of the study is presented in Section~\ref{sec-con}. Throughout this paper, we adopt units where $G=c=1$, unless otherwise specified.

\section{Method} \label{sec-met}
The ppE framework was originally introduced to investigate possible deviations from GR and probe alternative theories of gravity \citep{2009PhRvD..80l2003Y}. The ppE waveform model in inspiral regime is formulated as follows $\tilde{h}_{\rm ppE}(f)=\tilde{h}_{\rm GR}(f)(1+\alpha u^a)e^{i\beta u^b}$, where $(\alpha, a, \beta, b)$ are the ppE parameters. Here $u=\pi\mathcal{M}f$ is the inspiral reduced frequency, and $\mathcal{M}=(m_1m_2)^{3/5}/(m_1+m_2)^{1/5}$ is the chirp mass. The work by \citet{2019PhRvD.100j4001T} shows that sole phase corrections are capable of providing reliable constraints. Consequently, we focus on phase corrections arising from non-Kerr effects. The ppE waveform model can be simplified as 
$\tilde{h}(f)=A(f)e^{i\Psi_{\rm GR}(f)+i\beta_{\rm EdGB,charge}(\pi\mathcal{M}f)^b}$,
where the additional modifications to the GW phase appear at -1PN for both EdGB gravity and charge effect of BHs. Specifically, for EdGB gravity, the ppE parameters are \citep{2012PhRvD..85f4022Y}
\begin{eqnarray}
    \beta_{\rm EdGB} = -\frac{5}{7168}\frac{(m_1^2s_2^{\rm EdGB}-m_2^2s_1^{\rm EdGB})^2}{\eta^{18/5}M^4}\frac{16\pi\alpha_{\rm EdGB}^2}{M^4} ,
\end{eqnarray}
where, $\eta=m_1m_2/M^2$ is the symmetric mass ratio, $M=m_1+m_2$ is the total mass, and $s_i^{\rm EdGB}=2(\sqrt{1-\chi_i^2}-1+\chi_i^2)/\chi_i^2$ are the dimensionless BH charges in EdGB gravity. Here, $\chi_i=\vec{S_i}\cdot\hat{L}/m_i^2$ represents the dimensionless spins component of the BHs in the direction of the orbital angular momentum.
In NSBH binary system, the scalar charges $s^{\rm EdGB}$ vanish for neutron stars \citep{2016PhRvD..93b4010Y}. Considering that the phase corrections stemming from EdGB modifications are small perturbations compared to the GR contributions, the coupling parameters must satisfy the small coupling approximation $\zeta_{\rm EdGB}=16\pi\alpha^2_{\rm EdGB}/m_s^4<1$. Following Refs.~\citep{2023PhRvD.108d4061W, 2022PhRvD.105f4001L, 2021PhRvD.104b4060P}, we use the threshold
\begin{equation}
    \sqrt{\alpha_{\rm EdGB}} \le \frac{m_s}{(32\pi)^{1/4}} = 0.466\frac{m_s}{M_\odot} \mbox{[km]},
    \label{eq:limit}
\end{equation}
where $m_s$ is the smallest mass scale of a system and $M_\odot$ means solar mass.

In general, the waveform models of GW do not include the charge of BHs. By considering the charge effects as a perturbation in the inspiral stage, the corrections due to the electric dipole radiation on the waveform can be captured by the ppE framework. The coefficients can be written as \citep{Wang:2020ori, Christiansen:2020pnv, Cardoso:2016olt},
\begin{eqnarray}
    \Delta \Psi_{\rm charge} &&= \beta_{\rm charge}(\pi\mathcal{M}f)^b \nonumber\\
    &&= -\frac{5}{3584}\eta^{2/5}\zeta^2\kappa^2\left( \pi\kappa\mathcal{M}f \right)^{-7/3},
    \label{eq:charge}
\end{eqnarray}
where the difference between the charge of binary stars is denoted as $\zeta=|\lambda_1-\lambda_2|/\sqrt{1-\lambda_1\lambda_2}$, with $\lambda_i=q_i/m_i$ representing the charge-to-mass ratio and $q_i$ representing the electric charge. The above equation is applicable to binary systems where one or both components possess an electric charge. Although neutron stars are typically regarded as electrically neutral, several theoretical models propose the existence of charged neutron stars \citep{1975ApJ...196...51R,2005PhRvD..72j4017G,Fabbrichesi:2019ema}. Consequently, we focus on deriving constraints on the combination of electric charges of the binary components, rather than setting $\lambda_{\rm NS}$ to zero. In this work, the effects of charge are considered as small perturbations, i.e., $\kappa=1-\lambda_1\lambda_2 \approx 1$.

In the context of data processing, the posterior probability distribution of parameters $\vartheta$ given data $d$ is defined by the following expression:
\begin{equation}
    p(\vartheta|d,\mathcal{H}) = \frac{p(d|\vartheta,\mathcal{H})p(\vartheta|\mathcal{H})}{p(d|\mathcal{H})} = \frac{p(d|\vartheta,\mathcal{H})p(\vartheta|\mathcal{H})}{\int \mathrm{d}\vartheta p(d|\vartheta,\mathcal{H})p(\vartheta|\mathcal{H})} ,
\end{equation}
as derived from Bayes's theorem. Here, $p(d|\vartheta, \mathcal{H})$, $p(\vartheta|\mathcal{H})$ and $p(d|\mathcal{H})$ represent the likelihood, the prior, and the evidence, respectively. In cases of stationary and Gaussian noise, the likelihood is modeled as:
\begin{equation}
    p(d|\vartheta,\mathcal{H}) \propto \mathrm{exp}\left[-\frac{1}{2}\sum_{j=1}^N(d_j-h_j|d_j-h_j)\right] ,
\end{equation}
where $d_j$ and $h_j(\vartheta)$ denote the data and the waveform model for the $i$-th detector, respectively. The noise-weighted inner product $(a(t)|b(t))$ is defined by:
\begin{equation}
    (a(t)|b(t)) = 2\mathcal{R}\int_{f_{\rm low}}^{f_{\rm high}}\frac{\tilde{a}^*(f)\tilde{b}(f) + \tilde{a}(f)\tilde{b}^*(f)}{S_n(f)}\mathrm{d}f ,
\end{equation}
where $f_{\rm low}$ and $f_{\rm high}$ are the high-pass and low-pass cutoff frequencies of GW data, $*$ indicates the complex conjugate, and $S_n(f)$ is the power spectral density. In this study, we primarily examines the inspiral phase. According to Refs.~\citep{2016PhRvD..93d4007K,PhysRevD.100.104036}, the transition frequency from inspiral to intermediate phase is given by the relation $Mf_{\rm t}=0.018$, where $M$ is the detector frame's total mass of the binary. Therefore, we set the high-pass and low-pass cutoff frequencies at $f_{\rm low}=20$ Hz and $f_{\rm high}=0.018/[(m_1+m_2)]=584$ Hz, respectively. We conduct Bayesian parameter estimation for GW230529 using the BILBY package \citep{2019ApJS..241...27A} and the DYNESTY sampler \citep{Speagle:2019ivv} with 1000 live points. The number of sampling steps ranges from a minimum of 100 to a maximum of 5000. When bounding electric charges of the binary, we set $\kappa=1$ and sample over $\zeta$. Uniform priors are assigned for $\sqrt{\alpha_{\rm EdGB}}$ and $\zeta$ over the intervals $[0, 5]$ km and $[0, \sqrt{2}]$, respectively. For other parameters, we adhere to the priors utilized in Ref.~\citep{2024arXiv240404248T}.

\section{Results} \label{sec-res}

\begin{figure}
    \centering
    \includegraphics[width=0.45\textwidth]{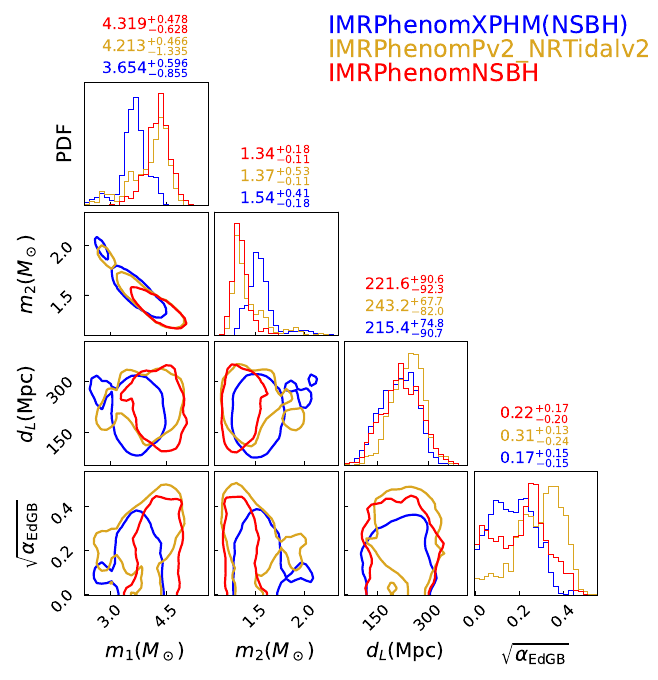}
    \caption{
    Posterior probability distributions of some key parameters, e.g., the gravitational coupling constant $\sqrt{\alpha_{\rm EdGB}}$, the masses of the primary ($m_1$) and secondary ($m_2$) components in the detector frame, and the luminosity distance ($d_L$) for GW230529. The golden, red, and blue colors represent analyses conducted with the IMRPhenomPv2\_NRTidalv2, IMRPhenomNSBH, and IMRPhenomXPHM (tidal effects are omitted) waveform models, respectively. The contours indicate the 90\% credible intervals. The diagonal entries display the median values and the corresponding 90\% interval ranges.}
    \label{fig:alpha_EdGB}
\end{figure}

\begin{figure}
    \centering
    \includegraphics[width=0.45\textwidth]{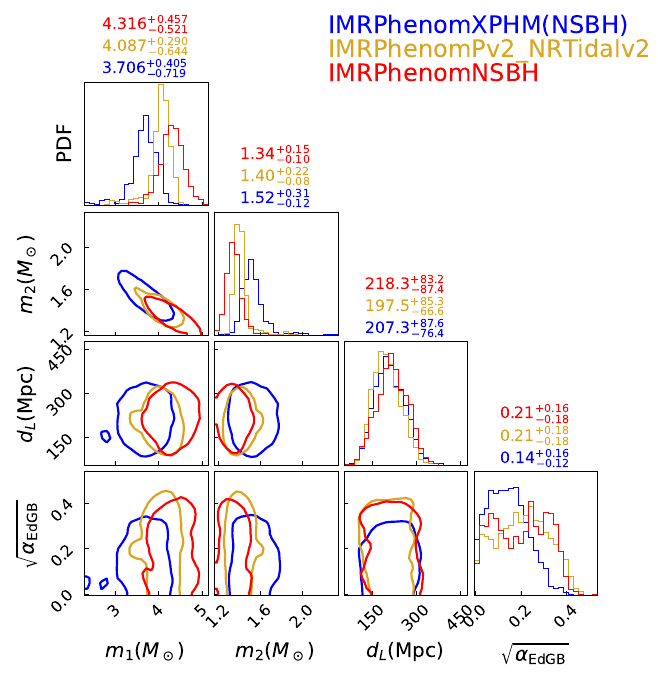}
    \caption{Similar to Figure~\ref{fig:alpha_EdGB} but for results with inclusion of high order corrections.}
    \label{fig:alpha_hpn_EdGB}
\end{figure}

To date, solar mass black holes have not been observed yet. GW230529 is more plausibly a NSBH merger \citep{Zhu:2024cvt, Chandra:2024ila}, although the possibility of a binary black hole (BBH) merger cannot be entirely ruled out based on sole GW data \citep{2024arXiv240405691H}. This study, therefore, investigates the constraints imposed by GW230529 on the interpretation of NSBH mergers. We present the posterior distributions derived from Bayesian analyses using various waveform models in Figure~\ref{fig:alpha_EdGB} and \ref{fig:alpha_hpn_EdGB}. The Figure~\ref{fig:alpha_EdGB} displays results that include -1PN corrections, while the Figure~\ref{fig:alpha_hpn_EdGB} incorporates results with higher-order PN corrections.
We obtain the constraints under various waveform models, including IMRPhenomNSBH \citep{2020PhRvD.101l4059T} and IMRPhenomPv2\_NRTidalv2 \citep{2019PhRvD.100d4003D}. Given the large mass ratio of GW230529, the tidal effects of the secondary star on the waveform are minimal. Accordingly, it is feasible to use the IMRPhenomXPHM model \citep{2021PhRvD.103j4056P}, initially developed for BBH mergers, to analyze NSBH events. 
For IMRPhenomPv2\_NRTidalv2 and IMRPhenomNSBH waveforms, uncertainties related to the equation of state (EoS) of neutron stars are not considered. Specifically, the primary star’s tidal deformability is assumed to be $\Lambda_1=0$, whereas the secondary star’s tidal deformability, $\Lambda_2$, is determined using the BSK24 EoS \citep{2013PhRvC..88f1302G, 2018MNRAS.481.2994P, 2019PhRvC.100c5801P}, consistent with the approach in Ref.~\citep{2024arXiv240404248T}.

\begin{figure}
    \centering
    \includegraphics[width=0.45\textwidth]{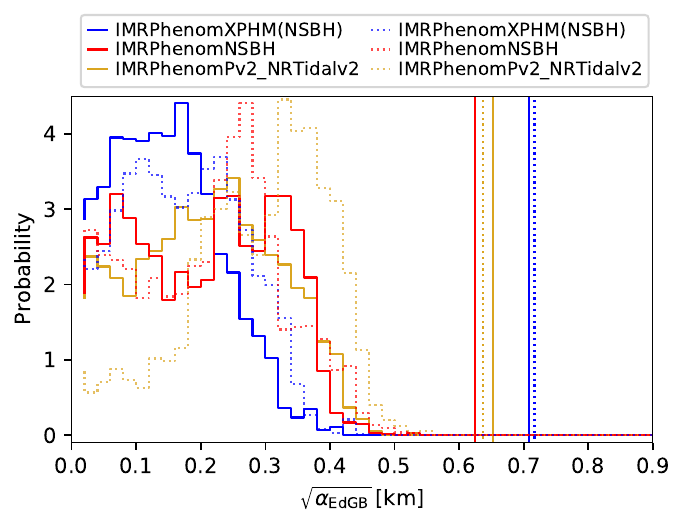}
    \caption{Comparison between the posterior distributions of $\sqrt{\alpha_{\rm EdGB}}$ and the requirements of the small coupling approximation. The vertical lines represent the thresholds of $\sqrt{\alpha_{\rm EdGB}}$ given by Eq.~(\ref{eq:limit}). It should be noted that the majority of the posterior distributions from each parameter estimation analysis fall within the threshold of the small coupling approximation. The solid and dotted lines denote results obtained with high order corrections and $-1$PN corrections, respectively.}
    \label{fig:limit}
\end{figure}

The constraints established by GW230529 on $\alpha_{\rm EdGB}$ are $\lesssim 0.415$ km for IMRPhenomPv2\_NRTidalv2, $\lesssim 0.361$ km for IMRPhenomNSBH, and $\lesssim 0.298$ km for IMRPhenomXPHM with tidal effects of the event neglected. Notably, IMRPhenomXPHM provides marginally better constraints on $\sqrt{\alpha_{\rm EdGB}}$ compared to other two waveform models, likely due to its more precise modeling.
For EdGB gravity, the leading corrections dominate at frequencies $f \lesssim 200$ Hz, whereas higher order corrections become significant at higher frequencies. Given the relatively low mass of the components in GW230529, the inspiral signals extend to higher frequencies, necessitating further investigation into $\alpha_{\rm EdGB}$ constraints in conjunction with advanced post-Newtonian (PN) phase corrections. The ppE framework, incorporating phase corrections up to 2PN order, is expressed as $\delta\Psi=\sum_{i}\beta_i^{\rm PPE}v^{-5+2i}$, where $v=(\pi Mf)^{1/3}$ and $i$ represents PN orders of $-1$, $0$, $0.5$, $1$, $1.5$, and $2$. The coefficients $\beta_i$ are detailed in Ref.~\citep{2022PhRvD.105f4001L}.
Incorporating higher-order corrections of EdGB gravity, the 90\% upper bounds are 0.260 km using the IMRPhenomXPHM model and 0.358 km using the IMRPhenomPv2\_NRTidalv2 model, representing improvements of 13\% and 14\%, respectively. As illustrated in Figure~\ref{fig:alpha_hpn_EdGB}, the median value and upper limit of $\sqrt{\alpha_{\rm EdGB}}$ have shown enhancements across all waveform models. We also check the validity of the small coupling approximation. As shown in Figure~\ref{fig:limit}, we find that GW230529 can place meaningful constraints on $\sqrt{\alpha_{\rm EdGB}}$ for all results. It is crucial to note that the higher PN corrections applied here do not account for tidal effects, which is outside the scope of this study. A summary of these bounds is provided in Table~\ref{tab:result_EdGB}.

\begin{table}[ht]
    \caption{The $90\%$ upper bounds on $\sqrt{\alpha_{\rm EdGB}}$ from GW230529} \label{tab:result_EdGB}
    \begin{ruledtabular}
    {
    \begin{tabular}{cccc}
    Waveform models & NSBH & NSBH (HPN) & Improvement \\\hline
    IMRPhenomXPHM & 0.298 & 0.260 & 13\%\\
    IMRPhenomPv2\_NRTidalv2 & 0.415 & 0.358 & 14\% \\
    IMRPhenomNSBH & 0.361 & 0.350 & 3\%
    \end{tabular}}
    \end{ruledtabular}
\end{table}

\begin{figure}
    \centering
    \includegraphics[width=0.45\textwidth]{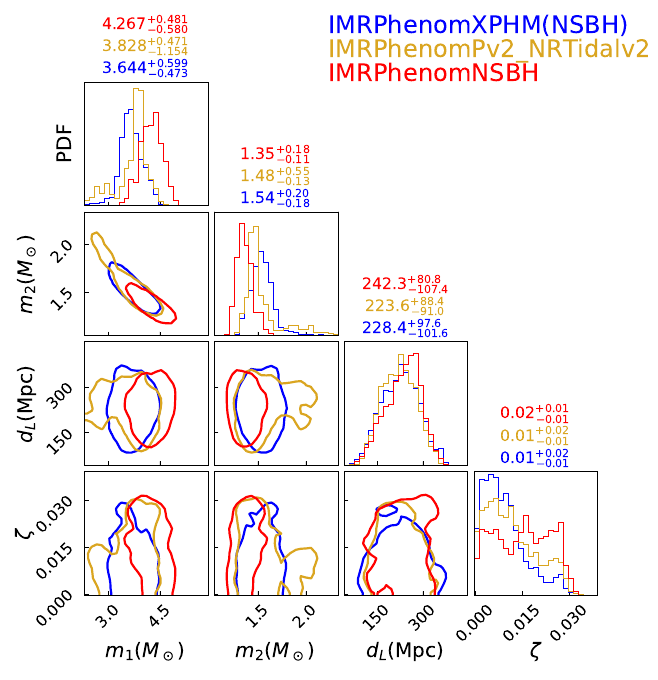}
    \caption{
    Posterior probability distributions of some key parameters, e.g., the combination of charges $\zeta$, the masses of the primary ($m_1$) and secondary ($m_2$) components in the detector frame, and the luminosity distance ($d_L$) for GW230529. The golden, red, and blue colors represent analyses conducted with the IMRPhenomPv2\_NRTidalv2, IMRPhenomNSBH, and IMRPhenomXPHM (tidal effects are omitted) waveform models, respectively. The contours indicate the 90\% credible intervals. The diagonal entries display the median values and the corresponding 90\% interval ranges.}
    \label{fig:charge}
\end{figure}

When investigating the potential charge of the binary, we do not consider non-GR gravity effects. Analysis conducted with the IMRPhenomPv2\_NRTidalv2 and IMRPhenomNSBH waveform models constrains the parameter $\zeta$ to be less than $0.024$ and $0.027$ at the $90\%$ credible upper level, respectively. When neglecting the tidal effects and utilizing the IMRPhenomXPHM waveform model, the constraint on the charge tightens to $\zeta \lesssim 0.019$. Figure~\ref{fig:charge} illustrates the posterior distribution of $\zeta$ for GW230529. These constraints are more stringent than the previously most restrictive limit of $0.21$ at the $90\%$ credible level, established by GW170608 \citep{Wang:2020ori}. The tightest bound on the charge-to-mass ratio of a single BH is $0.3$, achieved at $90\%$ credibility by the remnant BH of GW150914 \citep{2021PhRvL.126d1103B}. The constraints on the charge-to-mass ratio can be improved to $0.2$ with merger-ringdown signals of a GW150914-like event for the Einstein Telescope \cite{2024PhRvD.109b4058G}. Considering potential electromagnetic counterparts associated with GW200105, the estimated upper limit on the charge-to-mass ratio for the BH is $4.3 \times 10^{-3}$ \citep{2023PhRvD.108h3018Y}. Due to the unknown charge of NS in GW230529, the posterior distributions of the charge-to-mass ratio cannot be obtained from the distributions of $\zeta$. Nevertheless, even when the charge-to-mass ratio of an NS reaches as high as $0.1$\footnote{The charge of a uniformly magnetized NS is $Q=\frac{2}{3}\Omega B_{\rm P} R_{\rm NS}^3/c$ \citep{Ruderman:1975ju,2023PhRvD.108h3018Y}. For a $1.4 M_\odot$ NS with $B_{p}=10^{16}$ G, $P=1$ ms and $R_{\rm NS}=12$ km, the charge-to-mass ratio is $3.36\times10^{-3}$.}, the distribution of $\kappa$ (converted from the posterior of $\zeta$) is consistence with the small perturbation assumption ($\kappa \approx 1$) used in the parameter estimation. Consequently, the constraints on $\zeta$ provide meaningful limits.

\section{Conclusions} \label{sec-con}

In this work, we reanalyzed the GW event GW230529 within the ppE framework to impose constraints on the coupling parameters of EdGB gravity and the charge of this binary system. Given that GW230529 is a NSBH merger, we established an upper bound of $\sqrt{\alpha_{\rm EdGB}} \lesssim 0.298$ km at the 90\% credible level, with different waveform models resulting in slight variations in the constraints. By incorporating higher-order post-Newtonian phase corrections, we achieved tighter bounds, with $\sqrt{\alpha_{\rm EdGB}} \lesssim 0.260$ km. These are the first constraints derived from low mass NSBH mergers with higher-order post-Newtonian corrections. Considering the unknown nature of GW190814, the bounds presented here are the most stringent bounds among all robust results. The degree of improvement in these constraints varies across different waveform models. In addition to its implications for EdGB gravity, GW230529 also establishes a stringent limit on the charge of coalescing binary systems of compact objects, with an upper limit of 0.024 at the 90\% credible level. This is the first constraint from a NSBH system with GW data. While these constraints are less stringent than those derived from X-ray data, which estimate the charge-to-mass ratio of BHs to be on the order of $\lambda \sim 10^{-18}$, they nonetheless provide valuable insights obtained from GW observations.

\begin{acknowledgments}
The authors thank Yin-Jie Li, Ran Chen, and Wen-qing Guo for useful discussion. This work is supported in part by the National Natural Science Foundation of China under grant Nos. U2031205, 11733009, 12233011, 12303056, and 12247152, the China Postdoctoral Science Foundation (2022TQ0011), the Project for Special Research Assistant of the Chinese Academy of Sciences, and by the General Fund (No. 2023M733736) of the China Postdoctoral Science Foundation.

\end{acknowledgments}

\bibliographystyle{apsrev4-1}
\bibliography{re}

\end{document}